\documentclass[12pt]{article}
\usepackage{graphicx}
\usepackage{psfrag}

\def\beq{\begin{eqnarray}}
\def\eeq{\end{eqnarray}}
\def\m{M_*}
\def\msm{M_{\rm SM}}
\def\mpl{M_{\rm Pl}}
\def\e{{\epsilon}}

\def\lsim{\mathrel{\rlap{\lower3pt\hbox{\hskip0pt$\sim$}}
    \raise1pt\hbox{$<$}}}         
\def\gsim{\mathrel{\rlap{\lower4pt\hbox{\hskip1pt$\sim$}}
    \raise1pt\hbox{$>$}}}         

\begin{document}

\begin{flushright}
CERN --TH/2003-157\\
\end{flushright}

\vskip 1cm
\begin{center}
{\Large \bf
ICTP Lectures on Large Extra Dimensions
\vskip 0.2cm}
\vskip 1cm
{Gregory Gabadadze}
\vskip 1cm
{Theory Division, CERN, CH-1211  Geneva 23, Switzerland}\\
\end{center}

\vspace{0.9cm}
\begin{center}
{\bf Abstract}
\end{center}

I give a brief and elementary introduction to braneworld models
with large extra dimensions. Three conceptually distinct
scenarios are outlined: (i) Large compact extra dimensions; 
(ii) Warped extra dimensions;  (iii) Infinite-volume 
extra dimensions. As an example I discuss in detail 
an application of (iii) to late-time cosmology and 
the acceleration problem of the Universe.

\vspace{7cm}

\begin{center}
{\it Based on lectures given at: \\
Summer School on \\
Astroparticle Physics and Cosmology \\
Triese, Italy, June 17 -- July 5, 2002}\footnote{
A part of these lecture were also delivered at Fifth 
J.J. Giambiagi Winter School of Physics ``Precision Cosmology'',
July 28 -- August 1, 2003, Buenos Aires, Argentina.}

\end{center}

\vspace{0.1in}

\newpage

\vspace{4in}

\begin{center}
{\bf Disclaimer}
\end{center}

\vspace{0.1in}

Models with large extra dimensions have been studied
very actively during the last few years. 
There are thousands of works dedicated 
to the subject and any attempt of detailed account 
of those developments would require enormous efforts. 
The aim of the present work is to give a brief and  
elementary introduction to basic ideas and methods 
of the models with large extra dimensions and braneworlds.

The work is based on  lectures delivered at 
ICTP Summer School on Astroparticle Physics and Cosmology
for students with an introductory-level knowledge 
in classical and quantum fields, particle physics and cosmology.
The scope and extent of the lectures were restricted by the 
goals of the School. I apologize to those researchers who's 
advanced and original contributions to the subject 
could not be reflected in these lectures.

\newpage

\section{Introduction}  

The magnitude of gravitational force $F$ between two macroscopic objects 
separated at a distance $r$ obeys the inverse-square law,  
$F\sim r^{-2}$. This would not be 
so if the world had an $N\ge 1$ extra spatial dimensions 
that are similar to our three -- in that case 
we would instead measure $F\sim r^{-(2+N)}$. 
Similar arguments hold for micro-world of elementary particles.
For instance, we know from accelerator experiments that 
electromagnetic interactions of charged particles obey the 
inverse-square law.

However, experimental capabilities are limited and so is 
our knowledge of the validity of these laws of nature.
For instance, it has not been established how gravity behaves at 
distances shorter than $10^{-4}$ cm, or at distances 
larger than $10^{28}$ cm. All we know is that for 
$10^{-4} {\rm cm} \lsim r \lsim 10^{28} {\rm cm}$ the  
inverse square law provides a good description of nonrelativistic 
gravitational interactions, but laws of nature might be different
outside of that interval. Likewise, we are certain that 
electromagnetic interactions obey the inverse-square law 
all the way down to distances of order $10^{-16}$ cm, 
but they might change somewhere below that scale. 

At present it is not clear how  exactly these laws of nature
might change. There is a  possibility that they will change
according to the laws of higher-dimensional space if extra dimensions
exist. However, it is fair to wonder  why should one think 
in the first place that the world might have extra dimensions? 
I will give below major theoretical arguments that motivated 
an enormous amount of research in the field of extra dimensions.

The first scientific exploration of the idea of extra dimensions 
was by Kaluza \cite {Kaluza} and Klein \cite {Klein}. They
noticed that gravitational and electromagnetic interactions,
since so alike, could be descendants of a common origin.  
However, amazingly enough, the unified theory of gravity and 
electromagnetism was possible to formulate only in space 
with extra dimensions. Subsequently, non-Abelian gauge fields,
similar to those describing weak and strong interactions, 
were also unified with Einstein's gravity 
in models with extra dimensions. Therefore, 
the first reason why extra dimensions were studied was: 

\begin{itemize}

{\item  Unification of gravity and gauge interactions of 
elementary particles.}

\end{itemize}

So far we have been discussing classical gravitation.
However, quantization of gravity is a very nontrivial task. 
A candidate theory of quantum 
gravity, string theory (M-theory), can be formulated consistently in 
space with extra six or seven dimensions; Hence, 
the second reason to study extra dimensions:

\begin{itemize}

{\item  Quantization of gravitational interactions.}

\end{itemize}

All the extra dimensions considered  above were very small, 
of the planckian size and therefore undetectable. 
A new wave of activity in the field of extra dimensions came with 
the framework of Arkani-Hamed, Dimopoulos and Dvali (ADD) 
\cite {ADD} who observed that the Higgs mass hierarchy problem 
can be addressed in models with {\it large extra dimensions}.
Because the extra dimensions are large in  
the ADD framework,  their effects can be measurable 
in future accelerator, astrophysical 
and table-top experiments. Moreover, these models can be embedded 
in string theory framework \cite {AADD}. 
Subsequently Randall and Sundrum proposed a model with 
warped extra dimension \cite {RS2} that 
also provides an attractive setup for addressing 
the Higgs mass hierarchy  problem and for studying 
physical consequences of extra dimensions. Thus, the third reason is:

\begin{itemize}

{\item Higgs mass hierarchy problem.}

\end{itemize}

Another type of hierarchy problem is the problem of 
the cosmological constant. The latter is very hard to address
unless one of the conventional notions such as 
locality, unitarity, causality or four-dimensionality of space-time
is given up.  In that regard, theories with {\it infinite volume} 
extra dimensions \cite {DGP} -- the only theories  that are 
not four-dimensional at very low energies -- 
were proposed as a candidate for solving the cosmological 
constant problem \cite {DG,DGS}. Hence the fourth reason is:

\begin{itemize}

{\item Cosmological constant problem.}
 
\end{itemize} 

In what follows I will discuss some of the developments in 
extra dimensional theories listed above.

\section{Introduction to Kaluza-Klein Theories}

Extra spatial dimensions are not similar to our three dimensions
in the Kaluza-Klein (KK) approach. Instead, 
the extra dimensions form a {\it compact} space with 
certain compactification scale $L$. 
For instance, one extra dimension  
can be a circle of radius $L$, or simply an interval of size $L$.
For more than one extra dimensions this space could be a higher dimensional 
sphere, torus or some other manifold.
In general, $D$-dimensional space-time in the KK approach  
has a geometry of a direct product $M^4\times X^{D-4}$  
where $M^4$ denotes four-dimensional 
Minkowski space-time, and $X^{D-4}$ denotes a compact
manifold of extra dimensions --  called an {\it internal manifold}
\footnote{The $X^{D-4}$ does not have to be
a manifold in a strict mathematical definition of this notion 
(see examples below) however, we will use this name most of 
the time for simplicity.}. 

What is implied in the KK approach is that
there is a certain dynamics in $D$-dimensional space-time
that gives rise to preferential compactification of 
the extra $(D-4)$-dimensions leaving four minkowskian dimensions intact. 
The geometry $M^4\times X^{D-4}$ should be  
a solution of $D$-dimensional Einstein equations.

Let us now discuss what are the physical implications 
of the compact extra dimensions. 
Based on common sense it is clear that 
at distance scales much larger than $L$  the extra 
dimensions should not be noticeable. They only become ``visible'' 
when one probes very short distances of order $L$.

To discuss  these properties in detail we start with 
a simplest example of a real scalar field 
in $(4+1)$-dimensional space-time. In the the 
paper we use the mostly positive metric $[-++++..]$.
The Lagrangian density takes 
the form
\beq
{\cal L}\,=\,-{1\over 2}\,\partial_A\,\Phi \partial^A\,\Phi\,,
~~~~~A=0,1,2,3,{\it 5}\,. 
\label{lagr1}
\eeq
Here the field $\Phi(t,{\vec x},y)\equiv 
\Phi({x}_{\mu},y)\,, ~~{\mu =0,1,2,3}$, depends 
on four-dimensional coordinates $x_\mu$ as well as on 
an extra coordinate $y$. The extra dimension is assumed to be 
compactified on  a circle $S^1$ of radius $L$. Therefore, the 
five-dimensional space-time has a geometry of $M^4\times S^1$.
In this space the scalar field should be periodic with respect to 
$y \to y+2\pi L$:
\beq
\Phi({x},y)\,=\,\Phi({x},y+2\pi L )\,.
\label{period}
\eeq
Let us now expand this field in the harmonics on a circle
\beq
\Phi(x,y)\,=\,\sum_{n=-\infty}^{+\infty}\,\phi_n({x})\,e^{iny/L}\,.
\label{harmonics}
\eeq
(Note that $\phi^*_n({x})= \phi_{-n}({x})$).
Substituting this expansion into (\ref {lagr1})
the Lagrangian density (\ref {lagr1}) can be rewritten as
follows
\beq
{\cal L}\,=\,-{1\over 2}\,\,\sum_{n,m=-\infty}^{+\infty}\,
\left ( \partial_\mu \phi_n  \partial^\mu \phi_m\, -
\,{n m\over L^2}\,\phi_n \phi_m \right )e^{i(n+m)y/L}\,,
\label{lagr2}
\eeq
while the action takes the form
\beq
S\,=\,\int d^4x\int_0^{2\pi L}dy {\cal L}\,=\,-{2\pi L\over 2}\,
\int d^4x \, \sum_{n=-\infty}^{+\infty}
\left (\partial_\mu \phi_n  \partial^\mu \phi^*_n\,
+\,{n^2\over L^2} \phi_n \phi^*_n \right )\,.
\label{action}
\eeq
On the right hand side of the above equation we performed integration 
w.r.t. $y$. The resulting expression  is 
an  action for an infinite number of four-dimensional 
fields $\phi_{n}({x})$. To study properties of these fields 
it is convenient to introduce the notation
\beq
\varphi_n \equiv \sqrt{2\pi L} \phi_n \,. 
\label{varphi}
\eeq
The latter allows to rewrite the action in the following form
\beq
S\,=\,\int d^4x \, \left [ -{1\over 2}\,
\partial_\mu \varphi_0  \partial^\mu \varphi_0\  \right ]
\,- \,
\int d^4x \, \sum_{k=1}^{+\infty} 
\left (\partial_\mu \phi_k \partial^\mu \phi^*_k\,
+\,{k^2\over L^2} \phi_k \phi^*_k \right )
\label{actionspectrum}
\eeq
Therefore, the spectrum of a compactified theory consists of: 

\begin{itemize}

{\item A single real massless scalar field, called a {\it zero-mode},
$\varphi_0$;}

{\item An infinite number of massive complex 
scalar fields with masses inversely proportional to 
the compactification radius, $m_k^2=k^2/L^2$.}

\end{itemize}
All the states mentioned above  are called the Kaluza-Klein modes.
At low energies, i.e., when $E\ll 1/L$ only the zero mode
is important; while at higher energies $E\gsim 1/L$
all the KK modes become essential.

\vspace{0.2in}

As a next step we consider a $(4+1)$-dimensional example of
Abelian gauge fields.  An additional ingredient, compared to the 
scalar case, is the local gauge invariance the  consequences
of which we will emphasize below.

Let us start with the Lagrangian density
\beq
{\cal L}\,=\,-{1\over 4g_5^2}\,F_{AB}F^{AB}\,,
\label{lagrA}
\eeq
where the dimensionality's are set as follows:
$[A_B]=[{\it mass}]$, $ [g_5^{-1}]=[{\it mass}]$.
As in the previous example  we assume 
compactification on a circle $S^1$ of radius $L$ 
and periodic boundary conditions on the fields. 
We decompose $F^2_{AB}=F_{\mu\nu}^2+ 2 (\partial_\mu A_5 
-\partial_5 A_\mu)^2$, and expand the fields 
$A_\mu$ and $A_5$ in the harmonics on a circle
\beq
A_\mu(x,y)\,=\,\sum_{n=-\infty}^{+\infty}\,
A^{(n)}_\mu(x)\,e^{iny/L}\,,~~~~
A_5(x,y)\,= \,\sum_{n=-\infty}^{+\infty}\,
A^{(n)}_5(x)\,e^{iny/L}\,.
\label{harmonicA}
\eeq
As in the scalar example we integrate w.r.t. $y$ to calculate the 
effective 4d action
\beq
S\,=\,\int d^4x\int_0^{2\pi L}dy {\cal L}\,\equiv \,
\int d^4x\, {\cal L}_4 \,.
\label{actionA}
\eeq
Using gauge transformation the expression for ${\cal L}_4$ 
can be cast in the following
form
\beq
{\cal L}_4 \,=\,-{1\over 4g_4^2}\,\left \{ F^{(0)}_{\mu\nu}
F^{(0){\mu\nu}}\,+\,2 \sum_{k=1}^{+\infty}\left [ 
F^{(k)}_{\mu\nu}F^{*(k){\mu\nu}}\,+\,{2k^2\over L^2}\,
A^{(k)}_\mu A^{*(k)\mu}\,\right ]+\,2 (\partial_\mu A^{(0)}_5)^2
\right \}\,.
\label{L4}
\eeq
Therefore, we conclude that the spectrum of the compactified 
model consists of the following states:

\begin{itemize}

{\item A zero-mode --  a  massless gauge field $A_\mu^{(0)}$ 
with the gauge coupling $g_4^2=g_5^2/(2\pi L)$;}

{\item Massive KK gauge bosons
with the mass $m_k^2=k^2/L^2$;}

{\item Massless scalar field $A_5^{(0)}$. }

\end{itemize}

A few words on local gauge invariance are in order here.
The five-dimensional model is invariant under
five-dimensional local gauge transformations $A_B(x,y) \to
A_B(x,y) + \partial_B \alpha(x,y)$. 
After compactification  the five-dimensional 
gauge transformations  reduce  to an infinite number of 
{\it four-dimensional} gauge transformations --
one for each KK level $A_\mu^{(n)}(x) \to
A_\mu^{(n)}(x)  + \partial_\mu \alpha^{(n)}(x)$. 
However, only the zero-mode is massless gauge field, 
all the higher KK modes are massive.
This can be interpreted as a consequence  of the Higgs mechanism 
taking place on each massive KK level where 
a massless gauge field  ``eats'' one massless scalar $A_5^{(n)}$ 
and becomes a massive gauge field with 3 physical 
degrees of freedom. On the massless level there is a 
4d massless gauge field with 2 physical degrees of freedom plus 
one real massless scalar $A_5^{(0)}$.

\vspace{0.2in}

Finally we come to the main subject of this section and 
consider a $(4+1)$-dimensional example of gravity. 
It demonstrates how 4d Einstein gravity can be unified 
with electromagnetism in a 5d theory --- the original  proposal of 
Kaluza and Klein.

The 5d action takes the form
\beq
S={M_*^3\over 2}\,\int d^4x dy \sqrt {G}R_5\,.
\label{actionG}
\eeq
As in the previous examples the space is 
$M^{(4)}\times S^{1}$ and we expand fields in the harmonics 
on a circle of radius $L$
\beq
G_{AB}(x,y)\,=\,\sum_{n=-\infty}^{+\infty}\,
G^{(n)}_{AB}(x)\,e^{iny/L}\,.
\label{harmonicG}
\eeq
In what follows we will concentrate on 
the zero mode $G^{(0)}_{AB}$ neglecting all the massive modes.

Let us introduce the notations
\beq
G^{(0)}_{\mu\nu}=  e^{\phi/\sqrt{3}}(g_{\mu\nu}(x)+
e^{-\sqrt{3}\phi}A_\mu A_\nu)\,, \\ \nonumber 
G^{(0)}_{\mu 5}= G^{(0)}_{5 \mu }=e^{-2\phi/\sqrt{3}}A_\mu \,, 
\\ \nonumber
G^{(0)}_{ 5 5} = e^{-2\phi/\sqrt{3}} \,.
\label{KKanzats}
\eeq
Using these expressions we find the 4d action 
for the zero mode fields 
\beq
S_{\rm zm}\,=\,M_*^3 \pi L \int d^4x \sqrt{g}\left (  
R_4(g)-{1\over 2}\partial_\mu \phi \partial^\mu \phi
-{1\over 4}e^{-\sqrt{3}\phi }F_{\mu\nu}^2 \right )\,.
\label{Szm}
\eeq
Recalling that the conventional 4D action for gravity 
has a form
\beq
{\mpl^2\over 2}\,\int d^4x \sqrt{g}R_4(g)\,,
\label{4D}
\eeq
we find that $\mpl^2 =M_*^3 2\pi L$. As a result, the Newton constant
$G_N=(8\pi \mpl^2)^{-1}$ can be related to the higher dimensional scale
and the compactification radius
\beq
G_N = {1\over 16\pi^2 M_*^3 L}\,.
\label{GN}
\eeq
The main result of the above discussion is that four-dimensional 
gauge and gravitational fields have a common origin
in five-dimensional gravitational field. 

Let us count physical degrees of freedom. A four-dimensional 
massless graviton has 2 physical degrees of freedom (pdf's); 
A four-dimensional massless gauge boson has also 2 pdf's, 
and a real scalar has 1 pdf. Total is 5 pdf's, in agreement 
with 5 pdf's of a massless five-dimensional graviton 
\footnote{In general, the total number 
of independent components of a rank 2 symmetric tensor in 
$D$-dimensions is $D(D+1)/2$, however, only $D(D-3)/2$  
of those correspond to physical degrees of freedom of a 
$D$-dimensional massless graviton; the remaining extra components 
are the redundancy of manifestly gauge and 
Lorentz invariant  description of the theory.}.

Let us now turn to the massive KK levels.
The analysis is similar to that of  gauge fields but 
more cumbersome. Nevertheless, the main results can be summarised 
as follows. There is a massive graviton  with the mass
$m_k^2 =k^2/L^2$ at each $k$'th level. 
These gravitons acquire masses via the Higgs mechanism -- 
one massless graviton (2 pdf's) ``eats'' 1 massless gauge 
boson (2 pdf's) and one real scalar ( 1pdf) -- 
this makes one massive 4D graviton that
has 5 pdf's. The massive gravitational KK modes are charged under 
the massless gauge field. The charges are determined as 
$q_k \sim k/L\mpl\sim m_n/\mpl$. At the linearized level 
gauge transformations do not mix with each other
different KK levels, however, this mixing shows up once the 
nonlinear interactions of gravitational theory are taken into 
account \cite {Nappi}.

\section{Introduction to Braneworlds}

The idea that our $(3+1)$-dimensional world could  
be realized as a 3d surface in higher dimensional space  
was actively discussed in the context of general relativity 
the 1960th and 1970th.  

A first particle physics application of  this idea was put forward  
by Rubakov and Shaposhnikov \cite {RubShap} and 
independently by Akama \cite {Akama}. 

In this section, following \cite {RubShap}, we consider a toy 
example of the braneworld where the main mechanism  of localisation 
can explicitly be worked out.

We start with a  scalar field in 5-dimensions
with the following Lagrangian density
\beq
{\cal L}\,=\,-{1\over 2}\,\partial_A\,\Phi \partial^A\,\Phi\,-
{\lambda\over 2}\,\left (\Phi^2 - \eta^3   \right )^2\,.
\label{scalarL}
\eeq
The Lagrangian is invariant under the ${\bf Z}_2$ transformations
$ \Phi \to -\Phi$, however, the vacua of the theory are not ---
under the ${\bf Z}_2$ the two vacua $\Phi=\pm \eta^{3/2}$
interchange. Therefore, the ${\bf Z}_2$ is spontaneously broken.
As a result, there should exist domain walls.
We find the following domain wall (kink) 
solution to the classical equation of motion
\beq
\Phi_{\rm cl}(y)\,=\,\eta^{3/2}{\rm tanh}\left (\sqrt{\lambda}
\eta^{3/2}\,y\right )\,\equiv \,
\eta^{3/2}{\rm tanh}\left (m_0\,y\right )\,.
\label{kink}
\eeq
Transverse to the domain wall space is one dimensional,
hence, domain wall is a codimension one object.
Its worldvolume has 3 spatial coordinates, 
therefore, it is also called a 3-brane.

Let us discuss certain properties of the solution.
The tension of the wall is its  surface energy density
$T=\int dy H(\Phi_{\rm cl})=\int dy T_{00}(\Phi_{\rm cl})$,
where $H$ denotes the Hamiltonian and $T_{00}$ denotes the 
$00$ component of the stress tensor. The tension is determined as follows
\beq
T\,\sim\,{m_0^3\over \lambda}\,\sim\,\sqrt{\lambda}\eta^{3/2}\eta^3\,.
\label{tension}
\eeq

Below we would like to understand what are the excitations that 
live on the brane worldvolume. According to the 
braneworld idea \cite {RubShap},\cite {Akama}, 
in a realistic construction, those excitations should be 
identified with the Standard Model particles.
For this purpose we perform the following decomposition
\beq
\Phi(x,y)\,=\,\Phi_{\rm cl}(y)\,+\,\delta\Phi(x,y)\,.
\label{decomp}
\eeq
Then we find that the 5d equations have a solution 
\beq
\delta\Phi(x,y)\,=\,\left ({d\Phi_{\rm cl}\over dy}   \right )\,
\rho(x)\,,
\label{rho}
\eeq
where the four-dimensional field $\rho$ satisfies the 
equation
\beq
\partial_\mu^2\rho =0\,.
\label{eqrho}
\eeq
Therefore,  $\rho $ is nothing but a massless four-dimensional 
mode.  The wavefunction of this mode is proportional to 
${d\Phi_{\rm cl}/ dy}$ and vanishes outside of the brane.
Therefore, this mode is localized on a brane.
This excitation is just a Nambu-Goldstone boson of spontaneously
broken translation invariance along the $y$ direction.

\vspace{0.2in}

Let us now introduce fermions. For this we add to the Lagrangian
the following two terms
\beq
\Delta {\cal L}\,=\,i{\bar \Psi}\Gamma^M\partial_M\Psi -h \Phi 
{\bar \Psi} \Psi \,,
\label{psilagr}
\eeq
where $\Psi$ denotes a 5-dimensional Dirac fermion.
The equation of motion for the fermion in the background 
of the domain wall reads as follows:
\beq
i\Gamma^M\partial_M\Psi - h \Phi_{\rm cl}\Psi=0\,.
\label{eqpsi}
\eeq
This equation has a normalizable solution of the following form
\beq
\Psi_{\rm zm}(x,y)\,=\,e^{-\int_0^y h \Phi_{\rm cl}(z)dz}
\chi_L(x)\,,
\label{zm}
\eeq
where $\chi_L$ denotes a four-dimensional massless chiral mode
\beq
i \Gamma^\mu\partial_\mu \chi_L=0, ~~~~~\chi_L =(1-\gamma_5)\chi/2\,.
\label{chiL}
\eeq
From this expression we see that the wavefunction
of this mode vanishes outside of the brane.
Therefore, one obtains a four-dimensional chiral mode that is 
localized on the worldvolume\footnote{There also exists 
a solution with an opposite chirality that is 
not localized on a brane.}.

Summarizing, in a simple construction described above
scalars and fermions can be localized on a brane.
However, for realistic model building one should
in addition perform two major steps:

(i) Localize gauge fields on a brane;

(ii) Obtain four-dimensional gravity on the brane.

A mechanism for gauge field localisation
within the field theory context 
was proposed by Dvali and Shifman \cite {DS}.
It is based on the observation that gauge field can be in the 
confining phase the bulk while being in the broken phase on a brane; 
then confining potential prevents the low energy 
brane gauge fields to propagate into the bulk.
This mechanism is discussed in details in Refs. \cite {DS}.

Localisation of gauge fields is a rather natural property of 
D-branes in closed string theories \cite {Polchinski} -- 
the gauge fields emerge on a brane as 
fluctuations of open strings that are attached to the brane 
and do not exist in the bulk.

As to the issue (ii), below we discuss three distinct 
mechanisms by mens of which the laws of 4d gravity 
can be obtained on a brane.

\section{Braneworlds with Compact Extra Dimensions}

One way to obtain 4d gravity on a brane is 
to combine the braneworld idea with the 
idea of KK compactification. This, as was proposed
by Arkani-Hamed, Dimopoulos and Dvali (ADD) \cite {ADD},
opens up new possibilities to solve the Higgs mass hierarchy problem
and gives rise to new predictions that can be tested in 
accelerator, astrophysical and table-top experiments.  
Moreover, the framework can be embedded in string 
theory \cite {AADD}.

The main ingredients of a simplest ADD scenario are:

\begin{itemize}

\item{Standard Model particles are localized on a 3-brane,
while gravity spreads to all $4+N$ dimensions.}

\item{The fundamental scale of gravity $\m$, 
and the ultraviolet (UV) scale of the 
Standard Model, are around a few TeV or so.
This can eliminate the Higgs mass hierarchy problem.}

\item{$N$ extra dimensions are compactified.}

\end{itemize}
The action for a simplest ADD model takes the form:
\beq
S_{\rm ADD}={M_*^{2+N}\over 2}\,\int d^4x \int_0^{2\pi L}d^Ny
\sqrt{G}R_{(4+N)} + \int d^4x \sqrt{g}(T + 
{\cal L}_{\rm SM}(\Psi, M_{\rm SM}))\,
\label{SADD}
\eeq
where $M_*\sim (1-10)~{\rm TeV}$, $g(x)=G(x,y=0)$,
$T+\langle {\cal L}_{\rm SM} \rangle =0 $, the latter condition is a
usual fine-tuning of the cosmological constant. 

Technical simplifications which  are adopted above but that 
can be easily lifted are as follow:

(1) The Brane width is taken to be zero
(generically, the natural scale for the brane 
width could be  $M_*^{-1}$.) 

(2) Brane fluctuations are neglected (these are Nambu-Goldstone bosons
which couple to matter derivatively).

(3) All extra dimensions have equal size $L$
(in general, different extra dimensions could have different 
sizes).

(4) Only gravity can propagate in the bulk 
(in general, other fields could  also live in the bulk, in fact there 
are attractive scenarios with right-handed Neutrino living in the 
bulk \cite {Smirnov}.)

Let us first study the properties of 4d gravity in the 
ADD scenario. The low effective 4d action
for a zero mode takes the form
\beq
{M_*^{2+N}\over 2}\,\int d^4x \int_0^{2\pi L}d^Ny
\sqrt{G}R_{(4+N)} \rightarrow {M_*^{2+N} (2\pi L)^N\over 2}
\int d^4x \sqrt{g_{\rm zm}}R_{\rm zm}\,,
\label{zmgravity}
\eeq
hence, we should define the 4d Planck mass
\beq
\mpl^2 = M_*^{2+N} (2\pi L)^N \,.
\label{mplvsmstar}
\eeq
Postulating that the quantum gravity scale  
is  at $M_* \sim $TeV we find what should be 
the size of extra dimensions
\beq
L\sim 10^{-17+30/N}\,{\rm cm}\,.
\label{L}
\eeq
For one extra dimension, $N=1$, one gets  
$ L\sim 10^{13}$ cm, this is excluded within the ADD
framework since gravity below $10^{13}$ would have been
higher dimensional. For $N=2$ we get $ L\sim 10^{-2}$ cm; 
this particular case is very interesting
since it predicts modification of the 4d laws of 
gravity at submillimiter distances -- the subject of 
active experimental studies.
For larger $N$ the value of $L$ should decrease; but even for $N=6$
$L$ is very large compared to $1/\mpl$.

Two static sources on the brane interact with the 
following nonrelativistic gravitational potential
\beq
V(r)= -G_N m_1m_2 \sum_{n=-\infty}^{+\infty}|\Psi_n(y=0)|^2
{e^{-m_nr}\over r}\,,
\label{Vkk}
\eeq
where $\Psi_n(y=0)$ denotes the wavefunction 
of n'th KK mode at a position of a brane  and 
$m_n=|n|/L$. If $r\gg L$ from the above expression 
we find
\beq
V(r)=-{ G_N m_1m_2 \over r} \,.
\label{Vrlarge}
\eeq
This recovers the conventional 4d law of Newtonian dynamics.
In the opposite limit, i.e., when 
$r\ll L$ one gets
\beq
V(r)=-{ m_1m_2 \over M^{2+N}_* r^{1+N}} \,.
\label{Vshort}
\eeq
That is the law of $(4+N)$-dimensional gravitational interactions.
Therefore, the laws of gravity are modified at distances of order $L$. 

Selected topics of the ADD phenomenology:

\begin{itemize}

{\item  {\it Gauge coupling unification.} In a conventional 4d theory
the renormalization group running of the gauge coupling constants is 
logarithmic. This changes  in higher dimensions where the 
power-law  running takes place \cite {Veneziano}.
As was shown by Dienes, Dudas and Gherghetta  \cite {Keith},
the power-law running is what  gives rise to 
{\it an accelerated unification} of the strong, 
weak and electromagnetic couplings at a scale around
$\m$ in braneworlds with compact extra dimensions.}.

{\item {\it Missing energy signals in accelerator experiments}.
The SM particles are localized on a brane only up to some energy scale
that is comparable to $\m$. At about that  scale the SM particles
could in principle escape into the bulk. This would provide
missing energy signals in accelerator experiments. 
Another missing energy signal can be due to  emission of  
KK gravitons into the bulk, see detailed discussions in 
Refs. \cite {ADD},\cite{Wells}.}

{\item  {\it Energy loss by stars via emission of light KK gravitons}.
In the 6d ADD model  the KK gravitons are light,
$m_{KK}\sim L^{-1}\sim 10^{-4}{\rm eV}$. Therefore, these 
gravitons can be emitted in the interior of astrophysical  
objects the temperature of which exceeds $10^{-4}{\rm eV}$. 
As a result, these objects, such as stars, can coll down 
due to the process of emission of the KK gravitons into the bulk.
Each KK graviton emission if $\mpl$ suppressed.
However, because of the the high-multiplicity of KK graviton
the net result for the emission rate is suppressed by  $1/\m^2$.  
Unless this rate is small enough, 
a star would cool down faster than it should 
by  emitting  these KK gravitons. This puts a lower bout
on $\m$ in a 6d theory to be 50 TeV or so \cite {ADD,Per}.}

{\item  {\it Cosmological implications.}
There exist new  scenarios of inflation and Baryogenesis 
within the braneworld context. These scenarios 
manifestly use properties of branes. For instance,  
inflation on ``our brane'' can be obtained  if another brane falls 
on top of ``our brane'' in the early period of development of 
the brane-universe \cite {DvaliTye}. The potential that is created 
by another brane in ``our world'' can be viewed as the conventional 
inflationary potential. Baryon asymmetry of a desired magnitude 
can also be produced during the collision of these 
two branes \cite {DvaliGabad}. For more recent developments 
see Refs. \cite {E1},\cite {E2}, \cite {E3}, \cite {E4}.}

\end{itemize}

\section{Braneworlds with Warped Extra Dimensions  }

In this section we describe another way of obtaining 
4d gravity on a brane. It is based on a 
phenomenon of {\it localisation of gravity}
discovered by Randall and Sundrum (RS) \cite {RS2}.

We start with a so-called RS II model that has a 
single brane embedded in 5-dimensions bulk with negative cosmological 
constant. The action of the model is written as follows:
\beq
S_{\rm RS}={M_*^3\over 2}\,\int d^4x \int_{-\infty}^{+\infty}dy
\sqrt{G}( R_{5}- 2\Lambda) + 
\int d^4x \sqrt{g}(T + {\cal L}_{\rm SM}(\Psi, M_{\rm SM}))\,,
\label{SRS}
\eeq
where $\Lambda$ denotes the negative cosmological constant
and $T$ is the brane tension.

The equation of motion derived from this action 
takes the form (the Gibbons-Hawking surface term in the 
action is implied and hereafter we put $ {\cal L}_{\rm SM} =0$ 
for simplicity) 
\beq
\m \sqrt{G}\left ( R_{AB}-{1\over 2} G_{AB}R \right )=-\m^3
\Lambda \sqrt{G}G_{AB}+T\sqrt{g}\,g_{\mu\nu}\,\delta_A^\mu \,
\delta_B^\nu \,\delta(y)\,.
\label{eqRS}
\eeq
In our conventions the brane is located in extra space 
at the $y=0$ point. The above equations have a solution with a 
flat 4d worldvolume
\beq
ds^2 = e^{-|y|/L}\eta_{\mu\nu}dx^\mu dx^\nu +dy^2 \,,
\label{RSint}
\eeq
where $\eta_{\mu\nu}={\rm diag}(-+++)$ is the four-dimensional 
flat space metric, and we introduced the following notations
\beq
L\equiv \sqrt{ -{3\over 2\Lambda}}\,,~~~~~T={3\m^3\over L}\,.
\label{finetuning}
\eeq
The values of $\Lambda$ and $T$ 
have to be carefully adjusted to each other for this solution to exist.
Although the coordinate $y$ runs in the interval $(-\infty, +\infty)$
nevertheless, the physical size of extra dimension 
is finite:
\beq
\int_{-\infty}^{+\infty}dy\sqrt{G}\sim L\,.
\label{finitesize}
\eeq

The primary question that we would like to address is 
how does gravity look like on the brane?
For this let us consider graviton fluctuations:
\beq
ds^2 = \left ( e^{-|y|/L} \eta_{\mu\nu} + h_{\mu\nu}(x,y) \right )
dx^\mu dx^\nu +dy^2\,.
\label{perturb}
\eeq
We decompose $h_{\mu\nu}(x,y)\equiv u(y) {\tilde h}_{\mu\nu}(x)
= u(y) \e_{\mu\nu}{\rm exp}(ipx)$ with $p^2=-m^2$.
As a result the equation for the function $u$ takes the 
following form:
\beq
\left ( -m^2e^{|y|/L} -\partial_y^2 -{2\over L}\delta(y) +{1\over L^2}
\right ) u(y)=0\,.
\label{equ}
\eeq
For a zero-mode $m^2=0$ this equation 
simplifies and the solution can be found easily:
\beq
u(y)={\rm const.}e^{-|y|/L}\,.
\label{zmsol}
\eeq
Hence the interval for the zero-mode factorizes as follows
\beq
ds^2 = e^{-|y|/L} {\tilde g}_{\mu\nu}(x) dx^\mu dx^\nu +dy^2\,,
\label{intzmra}
\eeq
where we used the notations
${\tilde g}_{\mu\nu}(x)\equiv \eta_{\mu\nu} + {\tilde h}_{\mu\nu}(x)$.

It is important to emphasize that the five dimensional action 
is integrable w.r.t. $y$ for the zero-mode:
\beq
{M_*^{3}\over 2}\,\int d^4x \int_{-\infty}^{+\infty} dy
\sqrt{G}R_{(5)} \rightarrow {M_*^3(2 L)\over 2}
\int d^4x \sqrt{{\tilde g}}{\tilde R}\,.
\label{rszmcoupling}
\eeq
The result of this integration is a conventional 4d action. 
Hence we find a relation between the 4d Planck mass and $\m$
\beq
\mpl^2 =\m^3 (2L)
\label{mpmstarrs}
\eeq
This looks similar to the relation between the 
fundamental scale $\m$, the size of extra dimension
$L$ and the Planck mass $\mpl$ in the ADD model with one extra 
dimension. The similarity is due to the fact that 
the effective size of the extra dimension that is felt
my the zero-mode graviton is finite $\sim L$ as in the 
ADD as well as in the RS models. 

Besides the zero-mode there are  
an infinite number of KK modes \cite {RS2}.
Since the extra dimension is not compactified the KK modes 
have no mass gap. In the zero-mode approximation used in 
(\ref {rszmcoupling}) these states were neglected. 
However, at short distances $<< L$ the effects of those modes
become important. This can be seen by calculating
a static potential between sources on a brane.
The result reads:
\beq
V(r)=-{ G_N m_1m_2 \over r}\left (1+{(2L)^2\over r^2} 
\right ) \,.
\label{rspot}
\eeq
The second term in the parenthesis is due to the exchange of 
KK modes. We see that this term becomes dominant when $r\lsim L$.

\vspace{0.2in}

The above construction with the localized graviton can be used
for a new solution of the hierarchy problem.
This is achieved in a so-called RS I model \cite {RS1}.

The model contains two branes that are placed
at the endpoints of an interval of a certain size. 
One brane, called the ``hidden brane'',
has positive tension and 
the other one, called ``visible brane'', 
has negative tension. The equation of motion for 
this model looks as follows:
\beq
\m \sqrt{G}\left ( R_{AB}-{1\over 2} G_{AB}R \right )-  \m^3
\Lambda \sqrt{G}G_{AB} = \nonumber \\
T_{\rm hid}\, \sqrt{g_{\rm hid}}\,
g^{\rm hid}_{\mu\nu}\, \delta_A^\mu\, \delta_B^\nu \,\delta(y)\,+ \,
T_{\rm vis}\, \sqrt{g_{\rm vis}} \,
g^{\rm vis}_{\mu\nu}\, \delta_A^\mu \delta_B^\nu \, 
\delta(y-y_0)\,,
\label{rs1}
\eeq
were we used the notations
\beq
g_{\mu\nu}^{\rm hid}(x) = G_{\mu\nu}(x, y=0)\,,~~~
g_{\mu\nu}^{\rm vis}(x) = G_{\mu\nu}(x, y=y_0)\,.
\label{metricsrs1}
\eeq
As we mentioned above, the $y$ direction is compactified on an 
orbifold $S_1/{\bf Z}_2$ and $y$ runs in the interval $[-y_0,y_0]$.
One can check that there exists the following static solution to 
the equations of motion
\beq
ds^2 = e^{-|y|/L}\eta_{\mu\nu}
dx^\mu dx^\nu +dy^2\,.
\label{solrs1zm}
\eeq
The next step is find out fluctuations about this 
classical background. For this  
we proceed as in the RS II case. The derivation is straightforward
and the result is that the tensor 
$\eta _{\mu\nu}$ should be replaced as
$\eta _{\mu\nu} \to {\bar g}_{\mu\nu}(x)$, where  
\beq
g_{\mu\nu}^{\rm hid}(x) = {\bar g}_{\mu\nu}(x)\,,~~~
g_{\mu\nu}^{\rm vis}(x) = e^{-|y_0|/L} {\bar g}_{\mu\nu}(x)\,.
\label{metrichierarchy}
\eeq
Let us now look at what this leads to. For this we turn to the matter part 
of the Lagrangian. In the RS I it is assumed that the Standard Model fields 
are localized on a negative tension brane, i.e., at $y=y_0$.
As a representative SM field we consider the Higgs field $\phi$. 
We obtain:
\beq
\int d^4x \sqrt{g_{\rm vis}} \left \{ g^{\mu\nu}_{\rm vis}
(D_\mu \phi)^+(D_\nu \phi) -\lambda (|\phi|^2 - v_0^2)^2  
\right \} \to \nonumber \\
\int d^4x \sqrt{{\bar g}} \left \{ {\bar g}^{\mu\nu}
(D_\mu \phi)^+(D_\nu \phi) -\lambda (|\phi|^2 - e^{-y_0/L}v_0^2)^2  
\right \} \,.
\label{rshierarchy}
\eeq
Hence the Higgs VEV on a visible brane 
is rescaled by an exponential factor $v=e^{-y_0/2L}v_0$. 
Thus, all masses on the visible brane are suppressed by 
this exponential factor as compared to 
their natural values
\beq
m^2=e^{-y_0/L}m_0^2 \,.
\label{masssuppression}
\eeq
If $m_0\sim \mpl$, then in order to get $m\sim $TeV one needs
$y_0/L\sim 100$. Therefore, small hierarchy in $y_0/L$
gives rise to large hierarchy between $m$ and $m_0$.

The hierarchy problem is solved at the expense 
of fine tunning of the  tension of the hidden brane to the 
tension of the visible brane and both these tensions 
to the bulk cosmological constant. A possible way to avoid the fine 
tuning is to use the  stabilization mechanism proposed
by Goldberger and Wise \cite {GW}.
Another interesting scenario, studied by Karch and Randall \cite {KR} 
emerges when the tension and bulk 
cosmological constant are slightly detuned 
so that the worldvolume has $AdS_4$ geometry.
Regretfully, detailed discussion of these 
developments goes beyond the scope of the present lectures.

Selected topics of the RS phenomenology:

\begin{itemize}

{\item {\it Missing energy signals in accelerator experiments}.
The SM particles are localized on a brane only up to some energy scale
that is comparable with $\m$. At about that  scale the SM particles
could be emitted into the bulk. As in the ADD case, 
this would provide missing energy signals in accelerator experiments. 
See Ref. \cite {RSPhen} for details.}

{\item  {\it Gauge coupling unification.} In a conventional 4d theory
the renormalization group running of the gauge coupling constants is 
logarithmic. As we discussed before, 
this changes  in flat higher dimensions, the 
power-law  running takes place \cite {Veneziano}.
However, in the RS case the extra 5th dimension is not flat.
This affects dramatically the gauge coupling running
which can still be logarithmic as was discussed
in Refs. \cite {Pomarol}, \cite {Lisa}.}

\end{itemize}

\section{Braneworlds with Infinite Volume Extra Dimensions}

In this section we consider the third known mechanism  
of obtaining 4d gravity on a brane \cite {DGP}.
This mechanism is different from the previously 
discussed once since it allows the volume of the extra space to be 
infinite.
\beq
V_N\equiv \int d^Ny \sqrt {G}\to \infty \,.
\label{infvolume}
\eeq
The motivation for  constructing the
models with infinite-volume extra dimensions are as follow:

\begin{itemize}

\item{The size of extra dimensions does not need to be 
stabilized since the extra dimensions are nether compactified 
nor warped.}

\item{Because of the presence of infinite-volume 
extra dimensions gravity is modified at large distances.
This gives rise to  new solutions for late-time cosmology and 
acceleration of the universe.} 

\item{Although the supersymmetry should be broken
on a brane where the SM fields live, 
nevertheless  unbroken supersymmetry 
can be maintained in the bulk since it has an infinite volume}

\end{itemize}

The 5d model with these properties was proposed in \cite{DGP}
\beq
S_{\rm DGP}={M_*^3\over 2}\,\int d^4x \int_{-\infty}^{+\infty}dy
\sqrt{G}R_{5} + 
\int d^4x \sqrt{g}({\mpl^2 \over 2}R_4+ 
T + {\cal L}_{\rm SM}(\Psi, M_{\rm SM}))\,.
\label{actDGP}
\label{1}
\eeq
(The Gibbons-Hawking surface term is implied above). 
The main postulates in this approach are:

\begin{itemize}

\item{The SM fields are localized on a brane while gravity is propagating 
everywhere in extra dimensions}

\item{The UV cutoff of the SM $\msm \sim M_{\rm GUT} \gg \m$.}

\item{The brane width is assumed to be $1/\msm$.}

\item { $T+\langle {\cal L}_{\rm SM} \rangle =0 $, 
this is a usual fine tunning of the cosmological constant but 
the latter can be relaxed in higher co-dimensions \cite{DG,DGS},
in which case the model can be embedded into string theory 
\cite {Ignat}.}

\end{itemize}

What kind of gravity is described by this model?
Let us look at the gravitational part of (\ref {actDGP}).
We introduce the quantity
\beq
r_c \sim \mpl^2 /\m^3 \,,
\label{rc}
\eeq
When $r_c\to \infty$ the 4d term dominates, in 
the opposite limit $r_c\to 0$   the 5d term dominates.
Therefore we expect that for $r\ll r_c$ to recover
the 4d laws on the brane, while for $r\gg r_c$
5d laws. 

The 4d Ricci scalar $R_4=R_4(g(x))$ is constructed out of
the induced metric on a brane
\beq
g_{\mu\nu}(x)~\equiv~G_{\mu\nu} (x,~y=0)~.
\label{4dg}
\eeq
The Standard Model (SM) fields are confined to the brane. 
Note that the SM  cutoff
should not coincide in general with $\m $ and,  in fact, 
is assumed to be much higher in our case. For simplicity   
we suppress the Lagrangian of SM fields.
The braneworld origin of the action (\ref {1}) and parameters  $\m $, $\mpl$ 
were discussed in details in Refs. \cite {DGP,DG,DGKN}. 

Let us first study the non-relativistic potential between 
two sources confined to the brane.
For a time being we drop the tensorial structure in
the gravitational equations and discuss the distance dependence of the
potential. We comment on the tensorial structure below.

The static gravitational potential between the  
sources in the 4-dimensional
world-volume of the brane is determined as:
\beq
V(r)~=~\int G_R\left (t,{\overrightarrow x},y=0; 0,0,0\right
) dt~,
\label{pot}
\eeq
where $r\equiv\sqrt{x_1^2+x_2^2+x_3^2}$ and 
$G_R\left (t,{\overrightarrow x},y=0; 0,0,0\right )$
is the retarded Green's function (see below).
Let us turn to
Fourier-transformed quantities with respect to
the world-volume four-coordinates $x_\mu$:
\beq
G_R(x,y; 0,0)~\equiv~\int ~ {d^4p\over (2\pi)^4}~e^{ipx} ~{\tilde G}_R(p,y)~.
\label{Fourie}
\eeq
In  Euclidean momentum space the equation for the Green's function
takes the form:
\beq
\left (~ \m ^3(p^2-\partial_y^2)~+~\mpl^2~ p^2 ~\delta(y) ~\right )~
{\tilde G}_R(p,y)~=
~\delta(y)~.
\label{mom}
\eeq
Here $p^2$ denotes the square of an Euclidean four-momentum
$p^2\equiv p_4^2+p_1^2+p_2^2+p_3^2$.
The solution with appropriate boundary conditions
takes the form:
\beq
{\tilde G}_R(p,y)~=~{1\over \mpl^2p^2~+~2\m ^3p}~ {\rm exp} (-p|y|)~,
\label{sol1}
\eeq
where $p\equiv\sqrt {p^2}=\sqrt{p_4^2+p_1^2+p_2^2+p_3^2}$.
Using this expression and Eq. (\ref {pot}) one finds the following
(properly normalized)
formula for the  potential
\beq
V(r)~=~-{1\over 8\pi^2 \mpl^2}~{1 \over r}~\left \{ {\rm sin}
\left ( {r\over r_c} \right ) ~{\rm Ci} \left ( {r\over r_c} \right )
~+~{1\over 2}  {\rm cos}
\left ( {r\over r_c} \right ) \left
[\pi~-~2 ~ {\rm Si} \left ( {r\over r_c} \right ) \right ]   \right \}~,
\label{V}
\eeq
where
$ {\rm Ci}(z) \equiv \gamma +{\rm ln}(z) +\int_0^z ({\rm cos}(t) -1)dt/t$,
$ {\rm Si}(z)\equiv \int_0^z {\rm sin}(t)dt/t$,
$\gamma\simeq 0.577$  is  the Euler-Mascheroni
constant, and the distance scale $r_c$ is defined as follows:
\beq
r_c~\equiv ~{\mpl^2\over 2 \m ^3}~.
\label{r0}
\eeq
In our model we choose $r_c$ to be of the order of
the present Hubble size,  which is equivalent 
 to the choice $\m  \sim 10-100$ MeV. We will discuss 
 phenomenological compatibility
  of such a low quantum gravity scale below. 
It is useful to study  the short distance and long distance
behavior of this expression.

At short distances when $r<<r_c$ we find:
\beq
V(r)~\simeq~-{1\over 8\pi^2 \mpl^2}~{1 \over r}~\left \{
{\pi\over 2} +\left [-1+\gamma+{\rm ln}\left ( {r\over r_c} \right )
\right ]\left ( {r\over r_c} \right )~+~{\cal O}(r^2)
\right \}~.
\label{short}
\eeq
Therefore, at short distances the potential
has the correct 4d Newtonian $1/r$ scaling. 
This is subsequently modified
by the logarithmic {\it repulsion} term in (\ref {short}).

Let us turn now to the large distance behavior. Using (\ref {V})
we obtain for $r>>r_c$:
\beq
V(r)~\simeq~-{1\over 8\pi^2 \mpl^2}~{1 \over r}~\left \{
{r_c\over r}~+~{\cal O} \left ( {1\over r^2} \right )
\right \}~.
\label{long}
\eeq
Thus, the long distance potential
scales as $1/r^2$ in accordance with laws of 5d theory.

We would like to emphasize that the behavior (\ref {sol1})
is intrinsically higher-dimensional and 
is very hard to reproduced in conventional four-dimensional 
field theory. Indeed, the would be four-dimensional 
inverse propagator should contain the term $\sqrt{p^2}$.
In the position space this would correspond in the Lagrangian to  
the following pseudo-differential operator 
\beq
{\hat {\cal O}}~=~-~\partial_\mu^2~+~{\sqrt{-\partial_\mu^2}\over r_c}~.
\label{root}
\eeq
We are not aware 
of a consistent four-dimensional quantum field theory 
with a finite number of physical bosons which would lead to such 
an effective action. 

4D gravitational interactions in the present model are 
mediated by a resonance graviton with the 
lifetime $\tau \sim r_c$. The resonance-mediated gravity was 
first discussed in Refs. \cite {GRS,CEH,DGP0} in a different context. 
Yet another scenario in which the large distance gravity is 
modified due to the mass of a graviton was proposed and 
studied by the Oxford group \cite {IanKogan}.

Finally we would like to comment on the 
tensorial  structure of the graviton propagator in the 
present model. In flat space this structure is similar to that of
a massive 4d graviton \cite {DGP}. 
This points to the van Dam-Veltman-Zakharov 
(vDVZ) discontinuity \cite {Veltman,Zakharov}. 
However, this problem can in general be resolved by at least 
two methods. In the present context one has to use the results
of \cite {Vainshtein} where it was argued that 
the vDVZ discontinuity which emerges
in the lowest perturbative approximation is in fact absent in the 
full nonperturbative theory. The application of the similar arguments 
to  our model leads to the result which is continuous in $1/r_c$. 
This is discussed in details in Ref. \cite {disc}.
Thus, the vDVZ problem is an artifact of using 
the lowest perturbative approximation \footnote{
Note that the continuity in the graviton mass in (A)dS backgrounds was
demonstrated recently in Refs. \cite {Kogan,Massimo}. We should emphasize 
that we are discussing the continuity in the classical 
4d gravitational interactions on the brane.
There is certainly the discontinuity in the full theory in a sense that 
there are extra degrees of freedom in the model. These latter
can manifest themselves at quantum level in loop diagrams 
\cite {Duff}.}. 

In general, the simplest possibility to 
deal with the vDVZ problem, as was suggested in Ref.
\cite {DGKN}, is to compactify the extra space
at scales bigger than the Hubble size with 
$r_c$ being even bigger, 
but we do not consider this possibility here.

\vspace{0.2in}

{\it 1. Cosmological solutions}:
Below we  will mainly be interested in  
the geometry of the 4d brane-world and follow Ref. \cite {DGD}.  
For the completeness of the presentation
let us first recall the full  5d metric of the cosmological solution. 
The 5d line element  is taken in the following form: 
\beq
ds^2~=~-N^2(t,y)~dt^2~+~A^2(t,y)~\gamma_{ij}~dx^idx^j~+~B^2(t,y)~dy^2~,
\label{5dint}
\eeq
where $\gamma_{ij}$ is the metric of a 3 dimensional maximally 
symmetric Euclidean 
space, and the metric coefficients read \cite {Cedric}
\begin{eqnarray}
N(t,y)~&=&~1~+~\epsilon~|y|~{\ddot a}~({\dot a}^2~+~k)^{-1/2}~, \nonumber \\
A(t,y)~&=&~a~+~\epsilon~|y|~({\dot a}^2~+~k)^{1/2}~, \nonumber \\
B(t,y)~&=&~1~,
\label{nab}
\end{eqnarray}
where $a(t)$ is 4d scale factor and $\epsilon =\pm 1$.
Knowing the braneworld intrinsic geometry is all what matters
as far as 4d observers are concerned.
This geometry is given in the above solution. 
Taking the $y=0$ value of the metric we obtain  the 
usual 4d Friedmann-Lema\^{\i}tre-Robertson-Walker (FLRW) 
form (enabling to interpret $t$ as the cosmic time on the braneworld)
\begin{eqnarray} \label{FLRW}
ds^2~ &=&~ -dt^2~ + ~a^2(t)~ dx^i~ dx^j ~\gamma_{ij},\\
&=&  ~-dt^2~ + ~a^2(t)~\left(dr^2 + S^2_k(r)d\psi^2 \right),
\end{eqnarray}
where $d\psi^2$ is an  angular line element, $k=-1,0,1$ parametrizes
 the brane world  spatial curvature, and $S_k$ is given by 
\begin{eqnarray}
S_k(r) = \left\{\begin{array}{ll} \mbox{sin } r & (k=1) \\
\mbox{sinh } r &(k=-1) \\ r &(k=0) \end{array} \right \}
\end{eqnarray}
In the present case,
the dynamics is generically different from the 
usual FLRW cosmology, as shown in  \cite{Cedric}.
The standard first Friedmann equation is replaced in our model 
by 
\begin{equation} \label{fried}
H^2 + \frac{k}{a^2} = \left(\sqrt{\frac{\rho}{3 {\mpl^2}}  +
 \frac{1}
{4 r_c^2}} +\epsilon \frac{1}{2 r_c}\right)^2, 
\end{equation}
where $\rho$ is the total cosmic fluid energy density.
We have in addition the usual equation of conservation for 
the energy-momentum tensor of the cosmic fluid given by 
\beq \label{cons}
\dot{\rho} + 3H (p + \rho) = 0~.
\eeq
Equations (\ref{fried}) and (\ref{cons}) are sufficient 
to derive the cosmology of our model. In particular using 
these relations
one can obtain a second Friedmann equation as in 
standard cosmology.

Equation (\ref{fried}) with 
$\epsilon =1$ and $\rho =0$ has an interesting self-inflationary solution
with a Hubble parameter given by the 
inverse of the crossover scale $r_c$.
This can be easily understood looking back at the action  (\ref{1}) 
where it is apparent that the intrinsic curvature term on the brane 
appears as a source for the bulk gravity, so that with appropriate 
initial conditions, this term can cause  an expansion of the 
brane world without the need of matter or cosmological constant 
on the brane. This self-inflationary solution is the key ingredient for 
our model to produce late time accelerated expansion\footnote{
Note that the nonzero 4d Ricci scalar on the brane makes a seemingly 
 negative contribution to the 
brane tension \cite{Zurab,Cedric}. In this case, we 
consider a non fluctuating brane which is placed at the 
${\bf R}/{\bf Z}_2$ orbifold fixed point.}. Before discussing 
in detail this issue let us first  
compare our cosmology with the the standard one. 

We first note that the standard cosmological evolution is recovered 
from (\ref{fried}) whenever $\rho / {\mpl^2}$ is large  
compared to  $1/ r_c^2$, so that the early time cosmology of 
our model is analogous to standard cosmology. In this early phase
equation (\ref{fried}) reduces, at leading order,
to the standard 4d Friedmann equation given by 
\begin{equation} \label{stanfried}
H^2+ \frac{k}{a^2}~ = \frac{\rho}{3 {\mpl }^2}~.\end{equation}

The late time behavior is however generically different, as was 
shown in \cite{Cedric}: when the energy density decreases
and crosses the threshold $\mpl^2/r_c^2$,
 one either has a transition to a pure $5d$ regime 
(see e.g. \cite{bdl,bdel}) where the Hubble 
parameter is linear in
 the energy density $\rho$ (this happens for the $\epsilon =-1$ branch of the 
solutions), 
or to the self inflationary solution mentioned above (when $\epsilon =+1$).
This latter is the case we would like to investigate in more detail 
 in the rest of this work and we set $\epsilon =+1$ from now on. 
In terms of the Hubble radius (and for the flat Universe) 
the crossover between the 
two regimes happens when the Hubble radius $H^{-1}$
is of the order of the crossover length-scale between 
4d and 5d gravity,
that is $r_c$. 
If we do not want to spoil the successes of the ordinary cosmology, 
we have thus to assume the $r_c$ is of the order of the present 
Hubble scale $H_0^{-1}$. 

The conservation equation (\ref{cons}) is the same as the standard one, 
so that a given component of the cosmic fluid (non relativistic matter, 
radiation, cosmological constant...) 
will have the same dependence on  the scale factor as in standard cosmology. 
For instance, for a given component, labeled by $\alpha$, which has the 
equation of state $p_\alpha = w_\alpha \rho_\alpha$ (with $w_\alpha$ being a constant) one gets from  (\ref{cons})
$\rho_\alpha  = \rho_\alpha^0 a^{-3(1+ w_{\alpha})}$ 
(with $\rho_\alpha^0$ being a constant). The 
Friedmann equation 
(\ref{fried}) can be rewritten in term of the red-shift 
$1+z \equiv a_0/a$ as follows: 
\begin{equation} \label{H5d}
H^2(z) = H_0^2 \left\{ \Omega_k (1+z)^2 + 
\left( \sqrt{\Omega_{r_c}} + 
\sqrt{ \Omega_{r_c} + \sum_\alpha \Omega_\alpha(1+z)^{3(1+w_\alpha)} } 
\right)^2 \right\},
\end{equation}
where the sum is over all the components of the cosmic fluid.
 In the above equation  $\Omega_\alpha$ is defined as follows: 
\begin{equation} \label{omegafried}
\Omega_\alpha \equiv \frac{\rho^0_\alpha}{3 {\mpl}^2 
H_0^2 a_0^{3(1+w_\alpha)}}~,
\end{equation}
while $\Omega_k$ is given by 
\begin{equation}
 \Omega_k \equiv \frac{-k}{H_0^2 a_0^2}~,
\end{equation}
and $\Omega_{r_c}$ denotes  
\begin{equation}
\Omega_{r_c} \equiv \frac{1}{4 r_c^2 H_0^2}~.
\end{equation}
In the rest of this paper, as far as the cosmology of our model 
is concerned  we will consider a  
non-relativistic matter with density $\Omega_M$
in which case equation (\ref{H5d}) reads\footnote{Notice that we have set the cosmological constant 
on the brane to zero, 
and will do so until the end of this work since we are interested 
here in producing an accelerated Universes without cosmological constant.}.
\begin{equation} \label{HH5d}
H^2(z) = H_0^2 \left\{ \Omega_k (1+z)^2 + 
\left( \sqrt{\Omega_{r_c}} + 
\sqrt{ \Omega_{r_c} +  \Omega_M(1+z)^3 } 
\right)^2 \right\}.
\end{equation}
We can compare this equation with  the conventional 
Friedmann equation:
\begin{equation} \label{HH4d}
H^2(z) = H_0^2 \left\{ \Omega_k (1+z)^2 + 
\Omega_M(1+z)^3 +\Omega_X(1+z)^{3(1+w_X)}  \right\}~.
\end{equation}
Here, in addition to the matter and curvature contributions 
we have included the density of a dark energy component $\Omega_X$ 
with equation of state parameter $w_X$.
When $w_X=-1$, the dark energy acts in the same 
way as a cosmological constant, 
and the corresponding $\Omega_X$ will be denoted 
as $\Omega_\Lambda$ in the following. 
Comparing (\ref{HH5d}) and (\ref{HH4d}) we see that  
$\Omega_{r_c}$ acts similarly (but not identically, as we will see below) 
to a cosmological constant. 

\begin{figure}
\centering
\psfrag{a}{{\footnotesize open}}
\psfrag{b}{{\footnotesize closed}}
\psfrag{c}{{\footnotesize no big bang}}
\psfrag{z}{{\small $\Omega_M$}}
\psfrag{h}{\small {$\Omega_{r_c}$}}
\resizebox{12cm}{9cm}{\includegraphics{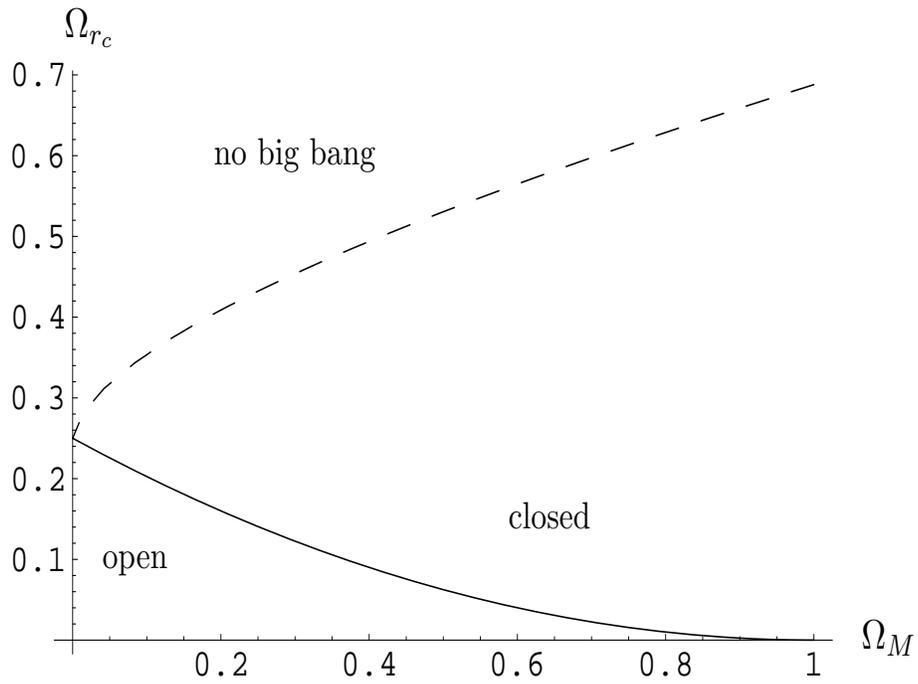}}
\caption{Different possibilities for the expansion 
as a function of $\Omega_M$ and $\Omega_{r_c}$. The solid line denotes a 
flat universe ($k=0$), with  $\Omega_{r_c}$ obtained through equation 
(\ref{flat5}). The Universes above the solid line are closed ($k=1$),
the universes below  are open  ($k=-1$). 
The Universes above the dashed line avoid the big bang singularity by 
bouncing in the past.}
\label{fig0}
\end{figure}

The $z=0$ value of equation  of equation(\ref{HH5d})
leads to the normalization condition: 
\begin{equation}\label{normalize}
 \Omega_k + \left( \sqrt{\Omega_{r_c}} + \sqrt{ \Omega_{r_c} +
  \Omega_M} \right)^2 =1,
\end{equation}
which differs from the conventional relation
\begin{equation}
\Omega_k  + \Omega_M + \Omega_X=1~.
\end{equation}  
For a flat Universe ($\Omega_k=0$) we get from equation 
(\ref{normalize}) 
\begin{equation} \label{flat5}
\Omega_{r_c} = \left(\frac{1-\Omega_M}{2}\right)^2 
\mbox{ and $\Omega_{r_c} <1$}.
\end{equation}
This shows in particular that for a  flat Universe, 
$\Omega_{r_c}$ is always smaller than $\Omega_X$, nevertheless, 
as will be seen below, the effects of  
$\Omega_{r_c}$ and  $\Omega_X$ can be quite similar. 
Figure \ref{fig0} shows the different possibilities 
for the expansion 
as a function of $\Omega_M$ and $\Omega_{r_c}$.

\vspace{0.2in}

{\it 2. Cosmological Tests}: We would like to discuss now, 
in a qualitative way, a few cosmological tests and measurements. 
We do not expect that the current experimental precision
would enable us to discriminate between the prediction of our  model 
and the ones of standard cosmology. However,  
the future measurements might enable to do so.

In order to  compare the outcome of our model with 
various cosmological tests  we need first to summarize  
some results. In the FLRW metric (\ref{FLRW}), we define, 
as usual (see e.g.\cite{Hogg:1999ad}),  
the transverse, $H_0$ independent (dimensionless),
comoving distance $d_M$: 
\begin{eqnarray}
\begin{array}{llll}
d_M &=& \frac{ S_k \left( \sqrt{|\Omega_k|} d_C \right)}{\sqrt{|\Omega_k|}},
&\mbox{ if $\Omega_k \neq 0$  }, \\
 d_M &=&  d_C ,& \mbox{ if $\Omega_k = 0$ },
\end{array}
\end{eqnarray}
where $d_C$ is defined as follows: 
\begin{equation}
d_C = \int_0^z H_0 \frac{dx}{H(x)}~.
\end{equation}
From the expression for $d_M$ one gets the ($H_0$ independent and
dimensionless) luminosity distance $d_L$ and the 
($H_0$ independent) angular diameter distance $d_A$ given by 
\begin{eqnarray}
d_L &=& (1+z) d_M, \\
d_A &=& \frac{d_M}{1+z}.
\end{eqnarray}
These definitions can be used  on the same footing both 
in standard and in our cosmological  scenarios (as they stand above,
they only rely on the geometry of the four-dimensional Universe
seen by the radiation  which is the same in both cases). 
The only difference is due to the expression for  $H(z)$ 
which enters the definition of $d_C$; one should choose either 
equation (\ref{HH4d}) or (\ref{HH5d}) depending on the case considered.  
Whenever we want to distinguish between the two models, we will put a 
{\it tilde} sign to the quantities corresponding to our model
(e.g. $\tilde{d}_L$).

\begin{figure}
\centering
\psfrag{z}{{\small z}}
\psfrag{h}{\small {$d_L(z)$}}
\resizebox{12cm}{7cm}{\includegraphics{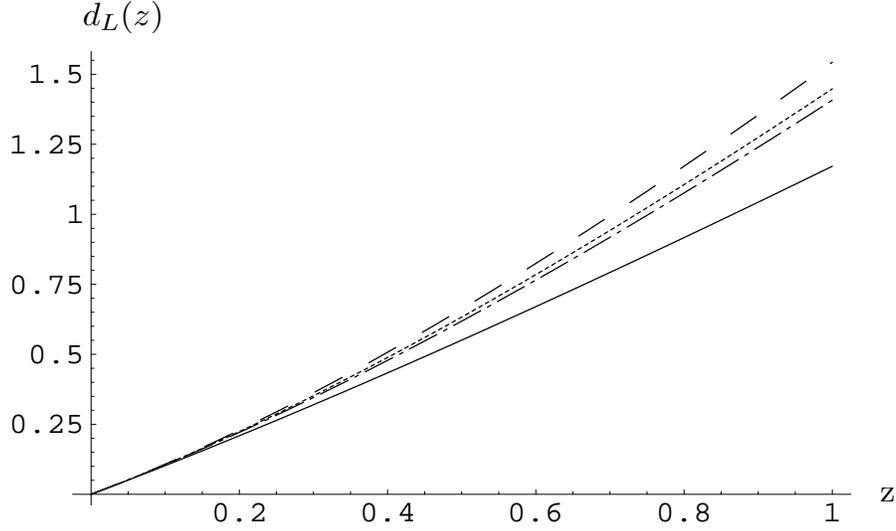}}
\caption{Luminosity distance as a function of red-shift
for ordinary cosmology with $\Omega_\Lambda=0.7, 
\Omega_M=0.3,k=0$ (dashed line),
$\Omega_\Lambda=0, \Omega_M=1,k=0$ (solid line),
and dark energy with $\Omega_X=0.7,w_X=-0.6, \Omega_M=0.3, k=0$ 
(dotted-dashed line) and in our model (dotted line) 
with $\Omega_M = 0.3$ and a flat universe 
(for which one gets from equation (\ref{flat5}) $\Omega_{r_c} = 0.12$
and $r_c =1.4 H_0^{-1}$).}
\label{fig2}
\end{figure}

\vspace{0.2in}{\it 2.1. Supernovae Observations}:
The evidence for an accelerated universe coming from 
supernovae observation  relies primarily 
on the measurement of the apparent 
magnitude of type Ia supernovae as a function of red-shift.
The apparent magnitude $m$ of a given supernova 
is  a function of its absolute magnitude $\cal M$, 
the Hubble constant $H_0$ and $d_L(z)$ 
(see e.g. \cite{Goliath:2001af}). 
Considering the supernovae as standard candles,  $\cal M$
is the same for all supernovae, so is $H_0$; thus,  
we need only to compare 
$d_L(z)$ in our model with that in standard cosmology.
Figure \ref{fig2} shows the luminosity distance $d_L$
as a function of red-shift in standard cosmology 
(for zero and non-zero cosmological constant) and in our model. 
This shows the expected behavior: 
our model mimics the cosmological constant in producing the late-time 
accelerated expansion. However, as is also apparent from this plot, 
for the same flat spatial geometry and the same amount of 
non-relativistic matter, our model does  not produce exactly the same 
acceleration as a standard cosmological constant,
but it rather mimics the one obtained from a dark energy 
component with $w_X > -1$.

\begin{figure}
\centering
\psfrag{z}{ {\small z}}
\psfrag{w04}{{\scriptsize $w_X=-0.4$}}
\psfrag{w06}{{\scriptsize $w_X=-0.6$}}
\psfrag{w08}{{\scriptsize $w_X=-0.8$}}
\psfrag{h}{\small {$d_L(z)/d_L^{ref}(z)$}}
\resizebox{12cm}{7cm}{\includegraphics{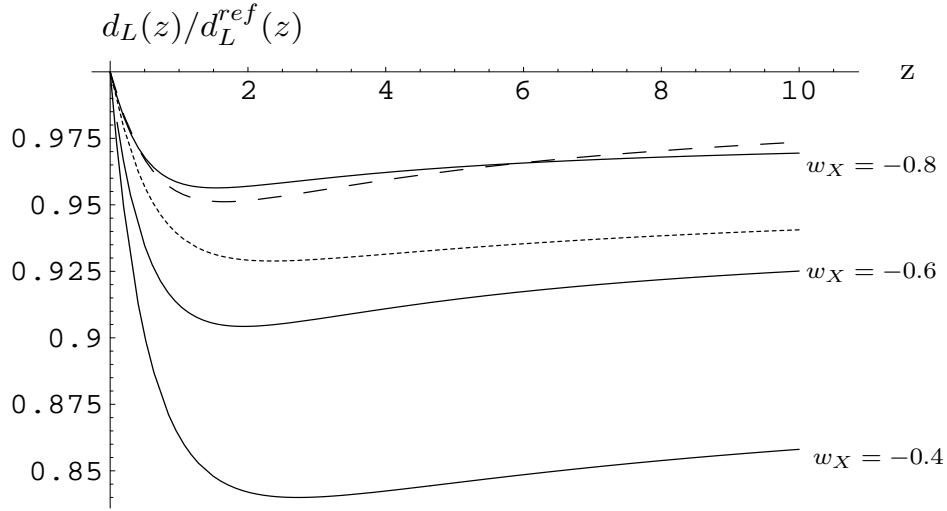}}
\caption{Plot of $d_L(z)/d_L^{ref}(z)$ 
for various models
of dark energy with constant equation of state parameters $w_X$ 
in standard cosmology (solid lines) as compared with the outcome of the  model
consider in this paper (dashed and dotted lines). 
All plots correspond to flat universes with $\Omega_M =0.3$ 
(solid lines, and dotted line), and  $\Omega_M =0.27$ (dashed line).}
\label{fig3}
\end{figure}

\begin{figure}
\centering
\psfrag{z}{ {\small z}}
\psfrag{w04}{{\scriptsize $w_X=-0.4$}}
\psfrag{w06}{{\scriptsize $w_X=-0.6$}}
\psfrag{w08}{{\scriptsize $w_X=-0.8$}}
\psfrag{h}{\small {$d_C(z)H(z)/H^{ref}(z) d_C^{ref}(z)$}}
\resizebox{12cm}{7cm}{\includegraphics{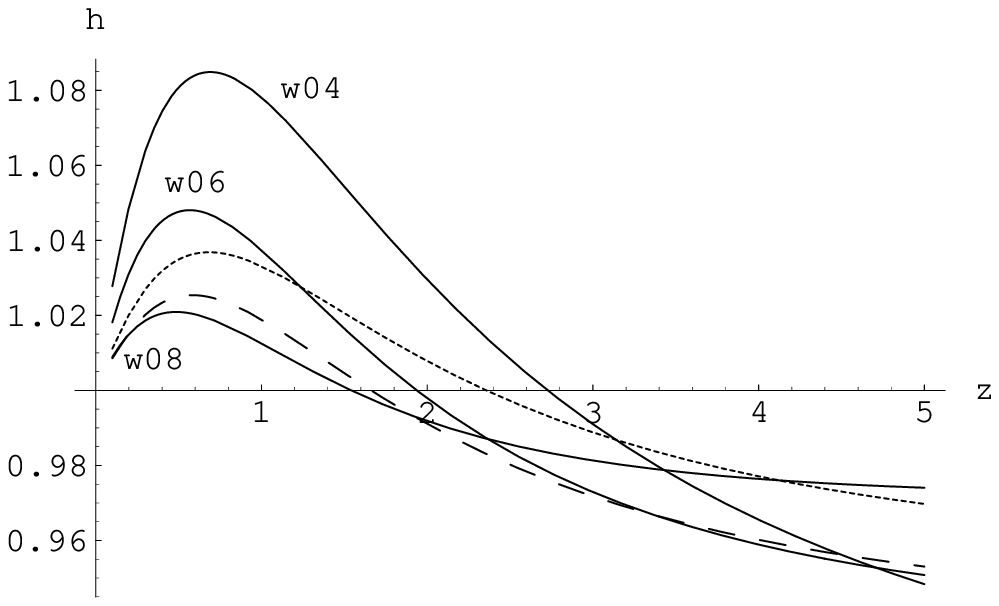}}
\caption{Plot of $H(z) d_C(z)/H^{ref}(z) d_C^{ref}(z)$ (Alcock-Paczynski test)
for various models
of dark energy with constant equation of state parameters $w_X$ 
in standard cosmology (solid lines) as compared with the outcome of the  model
considered in this paper (dashed and dotted lines). 
All plots correspond to flat universes, with $\Omega_M =0.3$ 
(solid lines, and dotted line), and  $\Omega_M =0.27$ (dashed line).}
\label{fig1}
\end{figure}

\vspace{0.2in}{\it 2.2. Comparison with dark energy}:
We want here to compare the predictions of our model to the ones of standard 
cosmology with a dark energy component. For this purpose we choose a reference 
standard model  given by standard cosmology  
with the parameters $\Omega_\Lambda =0.7$, $\Omega_M = 0.3$ and $k=0$ 
(and denote the associated quantities 
 with the superscript $^{ref}$, e.g. $d_L^{ref}$).
Figures \ref{fig3} and \ref{fig1} show respectively the 
luminosity distance $d_L(z)$ and $d_C(z) H(z)$ 
(Alcock-Paczynski test, see e.g.  \cite{Huterer:2000mj}) 
for various cases, showing that with precision tests, one should be 
able to discriminate between our model and a pure cosmological constant.

\begin{figure}
\centering
\psfrag{z}{ {\small $\Omega_M$}}
\psfrag{h}{\small {$\Omega_{r_c}$}}
\resizebox{10cm}{10cm}{\includegraphics{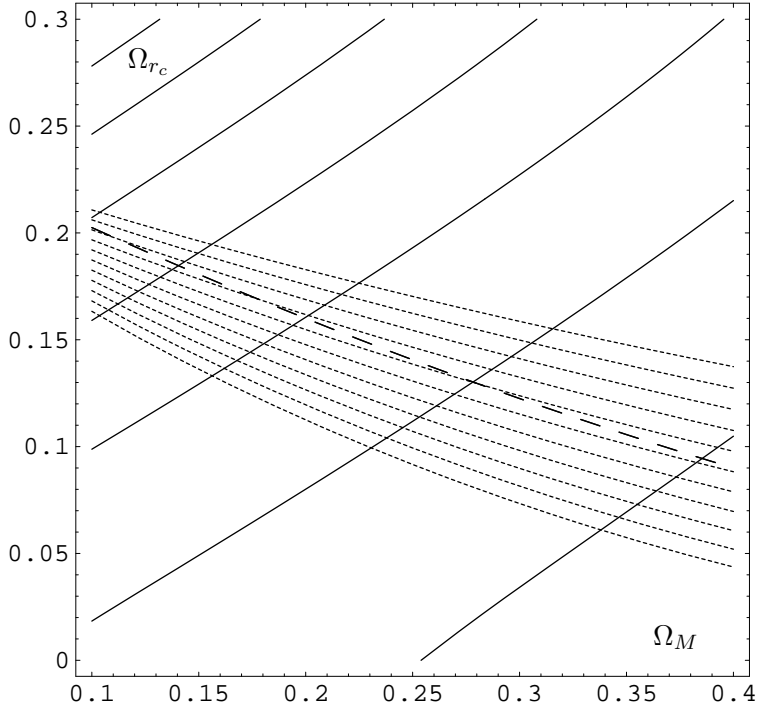}}
\caption{The Solid lines are lines
 of equal luminosity distance (in our model), 
$\tilde{d}_L(z=1)/d^{ref}_L(z=1)$, at red-shift 
$z=1$, the contours are drawn at every 5\% level. The dashed 
line corresponds to a flat universe. The dotted line are line of equal 
$\sqrt{\Omega_M}\tilde{d}_A(z)$ for $z=1100$, the contours are drawn 
at every 5\% level.} 
\label{fig5}
\end{figure}

\begin{figure}
\centering
\psfrag{z}{ {\small $\omega_{X}$}}
\psfrag{h}{\small {$d_A(z=1100)$}}
\resizebox{12cm}{7cm}{\includegraphics{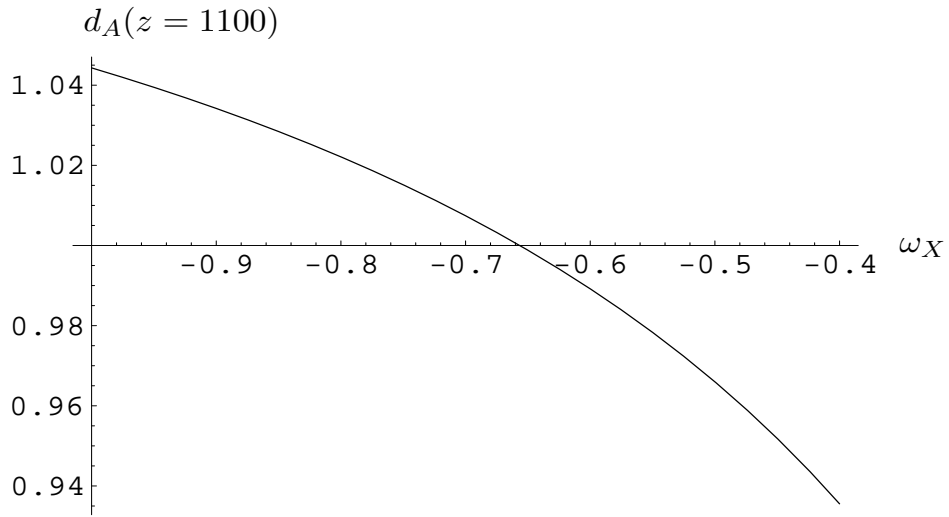}}
\caption{Angular diameter distance $d_A$ at $z=1100$ of standard cosmology
 divided  by $\tilde{d}_A(z=1100)$ in our model, as a function of $w_X$
 for a flat universe and $\Omega_M = 0.3$} 
\label{fig6}
\end{figure}

\vspace{0.2in} {\it 2.3. Cosmic Microwave Background (CMB)}:
It is well known that in standard cosmology, the 
location of points of constant 
luminosity distance at small $z$ 
is degenerated in the plane ($\Omega_M$, $\Omega_\Lambda$). This degeneracy 
can be lifted through CMB observations. Figure \ref{fig5} shows that this is 
the case as well in our model (Which should not be too much of a surprise, 
considering the similarities between early cosmology in the two models,
 as well as between the luminosity distances vs red-shift relations). 
The solid lines of figure \ref{fig5} are lines of constant $\tilde{d}_L$ 
at red-shift $z=1$; the dotted lines are lines of constant $\sqrt{\Omega_M} 
d_A$ at red-shift $z=1100$. This latter quantity roughly sets
 the position of the first acoustic peak in the CMB power spectrum,
 since its inverse measures the angular size on the sky of a physical length 
scale at last scattering proportional to $1/\sqrt{\Omega_M}$ 
(as is at first approximation the sound horizon at last scattering). 
Eventually figure \ref{fig6} shows the angular diameter 
distance $d_A$, at $z=1100$,  of standard cosmology divided by 
$\tilde{d}_A$ in our model, 
as a function of $w_X$, for a flat universe and $\Omega_M = 0.3$. 
This shows that, for the same content of matter (and a flat universe), 
the first Doppler peak 
in our model will be slightly on the left of the one obtained in 
standard cosmology with a pure cosmological constant.

One might wonder whether it is possible to obtain
the similar cosmological scenario in purely 
four-dimensional theory by introducing additional generally 
covariant  terms in the Einstein-Hilbert action.
The conventional local terms which can be added to 
the 4d theory contain higher derivatives. 
\beq
\mpl^2 ~\sqrt{g}~
\left (R~+~\alpha~{R^2\over \mpl^2}+...\right )~.
\label{hd}
\eeq
Whatever the origin of these terms might be their contributions should 
be suppressed
at distances bigger than  millimeter. 
That  is required by existing precision gravitational measurements. 
This implies  
that at distances of the present Hubble size their contributions are
even more suppressed. 
For instance, from the requirement  
that the contribution of the $R^2$ term 
to the Newtonian interaction  be  sub-dominant 
at distances around a centimeter implies that the relative 
contribution of the $R^2$ term at the Hubble scale is suppressed 
by the factor $({\rm cm}^2 H_0^2)~\sim~10^{-56}$. 
The contributions of other higher terms  
are suppressed even stronger.

It seems that the only way to accommodate this unusual behavior in 
a would-be pure 4d theory of gravity is to introduce terms 
with fractional powers of  the Ricci scalar,
for instance, such as  the term $\sqrt{g~R}$. However, it is hard to make sense of such a theory.

Therefore, we conclude that the scenario 
discussed in the previous sections is 
intrinsically high-dimensional one.

\vspace{0.2in} {\it 3. Constraints} In the preset  framework such a
low five-dimensional Planck scale is compatible with all the 
observations \cite {DGKN}.
In fact, at distances smaller that the present horizon size the 
brane observer effectively sees a single 4d 
graviton which is coupled with the strength 
$1/\mpl$ (instead of a 5d graviton coupled by 
the $1/\m ^{3/2}$ strength).

As it was shown in \cite{DGKN}  the high energy
processes place essentially no constraint on the scale $\m $.
This can be understood in two equivalent ways,
either directly in five-dimensional pictures, 
or in terms of the expansion in 4d  modes.

As was shown above, in five-dimensional language the brane 
observer at high energies sees graviton which is 
indistinguishable from the four-dimensional
one;  for short distances 
the propagator of this graviton is that of a 4d theory
\beq
{\tilde G}_R(p, y=0)~\propto~{1\over p^2}~.
\label{G5massless}
\eeq
Moreover, this state couples to matter 
with the $1/\mpl^2$ strength. 
Therefore in all the processes  with typical
momentum $p << 1/r_c$ the graviton production 
must go just like in 4d theory.
For instance, the rate of the graviton production 
in a process with energy $E$ scales  as
\begin{equation}
\Gamma ~\sim ~{E^3 \over \mpl^2}~.
\label{rate}
\end{equation}
The alternative language is that of the mode expansion. From the point of
view of the four-dimensional brane observer a single 5-dimensional
massless graviton is in fact a continuum of four-dimensional states,
with masses labeled by a
parameter $m$
\begin{equation}
G_{\mu\nu}(x,y) = \int dm~ \phi_m(y)~ h^{(m)}_{\mu\nu}(x)~.
\end{equation}
The crucial point is that the wave-functions of the massive modes
are suppressed on the brane as follows
\begin{equation}
|\phi_m(y=0)|^2~ \propto~{1\over 4~+ ~m^2~r_c^2}~.
\end{equation}
This is due to the intrinsic curvature term on the brane
which ``repels'' heavy modes off the brane \cite{DGKN,carena}. 
As a result their production in
high-energy processes on the brane is very difficult.
Let us once again consider bulk graviton production in a
process with  energy $E$ (e.g. star cooling via graviton emission at temperature $T$ of order $E$).
 This rate is given by \cite {DGKN}
\begin{equation}
 \Gamma ~\sim ~{E^3 \over \m ^3}~\int_0^{m_{\rm max}} ~dm ~
|\phi_m(0)|^2~.
\end{equation}
Here the integration goes over the continuum of bulk states  up to a maximum
possible mass which  can be produced in a given process $m_{\rm max} \sim E$.
However, since heavier wave-functions are suppressed on the brane by a
factor ${1 \over
m^2r_c^2}$, the integral is effectively cut-off at $m \sim 1/r_c$, which
gives for the rate
\begin{equation}
  \Gamma \sim {E^3 \over \m ^3~r_c} \sim {E^3 \over \mpl^2}~.
\end{equation}
This is in agreement with Eq. (\ref{rate}) and in fact
coincides with the rate of production of a single  
four-dimensional graviton, which is
totally negligible. Thus high-energy processes place no constraint on
scale $\m $ \cite {DGKN}. 

Due to the same reason cosmology places no bound on the scale $\m $.
Indeed, the potential danger would come from the fact that the early
Universe may cool via graviton emission in the bulk, which could affect
the expansion rate and cause deviation from an ordinary FLRW cosmology.
However, due to extraordinarily  suppressed graviton emission at high
temperature, the cooling rate due to this process is totally negligible.
Indeed in radiation-dominated era, the cooling rate due to graviton
emission is
\begin{equation}
\Gamma~ \sim~ {T^3\over \mpl^2}~.
\label{rad}
\end{equation}
At any temperature below $\mpl$ this is much smaller that the expansion
rate of the Universe $H ~\sim~ T^2/\mpl$. Thus essentially until $H \sim
\m ^3/\mpl^2$ (which only takes place in the present epoch) Universe
evolves as ``normal''.

The only constraint in such a case comes from the 
measurement of Newtonian force, which implies $\m ~> ~10^{-3}$eV
(this will be discussed in more detail elsewhere).

\vspace{0.2in} {\it 4. Dissipation}:
In the previous sections we established that 
classically the asymptotic form of the 4d metric on the brane is 
that of de Sitter space. 
Here we would like to ask the question whether this
asymptotic form can be modified due to quantum effects. 
This could happen if there is dissipation of the 
energy stored in the expectation value of the 4d Ricci scalar
into other forms which either can radiate into the bulk or be 
red-shifted away on the brane.  
Below we shall identify such a mechanism of potential 
dissipation.

An observer in de Sitter space is 
submerged in a thermal bath with nonzero temperature 
due to Hawking radiation from the de Sitter horizon.
The temperature of this radiation is $T\sim H$.
The crucial point is that 
the energy stored in this radiation can  
dissipate into the bulk in the form of vary long-wavelength 
graviton emission from the brane.
To estimate the rate of this dissipation we can use Eq. (\ref {rad})
with $T\sim H$. The corresponding change of the 
brane energy density in the absence of other forms of 
matter and radiation is given by:
\beq
{ d \rho_{\rm eff} \over  dt}~=~- {H^3\over \mpl^2}~ \rho_{\rm eff}~,
\label{rhoeff}
\eeq 
where $\rho_{\rm eff}~\equiv~\mpl^2~\langle R \rangle $ and 
the Hubble parameter can be written as $H^2~\propto~\langle R \rangle$.
The corresponding decay time is huge $\tau~\sim~10^{137}$ sec.
Therefore, the 4d metric eventually  asymptotes to flat Minkowski space.
Note the crucial difference from the conventional 4d de Sitter space
where the vacuum energy cannot dissipate anywhere due to the Hawking 
radiation.  In our case the existence of infinite volume
bulk is vital.

\vspace{0.2in} {\it 5. Infinite Volume and String Theory}
If the recent observations on the cosmological constant are
confirmed  it may be extremely  nontrivial to
describe the accelerated  Universe within
String Theory \cite {Banks,Witten}.
To briefly summarize the concerns
let us consider a generic theory with extra dimensions.
Usually one is looking for a ground state of the theory
with compactified or warped extra dimensions.
In both of these cases there is a length scale which defines
the volume of the extra space.  This scale cannot be bigger
than a millimeter \cite {ADD}.
Therefore, at larger distances 
a conventional 4-dimensional space is recovered.
Astrophysical observations indicate that this latter
asymptotes to the state of 4-dimensional accelerated expansion similar 
 to 4d de Sitter. In which case the following two
 problems may emerge \cite {Banks,Witten}:

\begin{itemize}

\item{An  observer in  dS space sees a finite portion of the space
bounded by event horizon.
In fact, the four-dimensional  dS interval  can be transformed
into the form:
\beq
ds^2_{\rm dS}~=~-\left (1~-~H^2 ~u^2 \right )~d\tau^2~+~
{du^2 \over (1~-~H^2 ~u^2)}~+~u^2~d\Omega_2~.
\label{ds}
\eeq
An observer is always inside of a finite size horizon.
As was argued in \cite {Banks}
physics for any such an observer is described by a finite number
of degrees of freedom\footnote{Indeed, the number 
of degrees of freedom inside
the region bounded by the horizon is finite.
Moreover, physics of the exterior of the horizon can in principle
be  encoded into the information on the horizon.
This latter, according to the Beckenstein-Hawking formula,
has finite entropy and, therefore,
supports a finite number of degrees of freedom.}.
On the other hand, there are an infinite number of 
degrees of freedom in String Theory and it is not obvious
how String Theory can be reconciled 
with this observation.}

\item{Another related difficulty is encountered
when on tries to define
the String Theory S-matrix on dS space.
As we mentioned above,
we  could think of dS space as
a cavity with a shell surrounding it.
This shell has nonzero temperature.
Thus, particles in the cavity are immersed in a thermal bath
and, moreover, there are no asymptotic states of free particles
required for the definition of the S-matrix.
It was shown recently that these problems generically persist
\cite {Sus,Fish} in quintessence models of
the accelerating Universe.}

\end{itemize}

Both of these difficulties  are related to the fact that
in dS space the comoving volume of the region which can be probed in the 
future by an observer is finite (the same discussion applies to  any
accelerating  Universe with $-1<w<-2/3$, where the
equation of state is $p=w\rho$).

The theories with infinite-volume extra dimensions
might evade these difficulties.
The reason is that the accelerating Universe in this case
can be accommodated in a space which is not simply 4-dimensional
dS.  In fact, as we argued in previous sections, although
the space on the brane looks like de Sitter space
for long time, it will asymptote to space with 
no dS horizon in the infinite future.
 
Let us discuss briefly these issues.

We start by  counting the  number of degrees of freedom
which are in contact with a braneworld  observer.
It is certainly true that an observer on the brane is bounded
in the world-volume dimensions
by a  dS horizon. However, there is no horizon in the
transverse to the brane direction. Thus, any observer on a brane
is in gravitational contact with
infinite space in the bulk. In this case,
the infinite number of bulk
modes of higher dimensional graviton participate in 4d interactions
on the brane \cite {DGP,DGKN}.
Therefore, the number of degrees of freedom
needed to describe physics on the brane is infinite.

The problem of definition of the S-matrix might be more subtle.
Below we present a simplest possibility.
The key observation is that
the metric (\ref {nab}) in the bulk is nothing but the
metric of flat Minkowski space.
Indeed, performing the following  coordinate transformation \cite {derul}:
\begin{eqnarray}
Y^0~&=&~A~\left ({r^2\over 4}~+~1~-~{1\over 4 {\dot a}^2}   \right )
-{1\over 2} \int dt ~{ {a}^2 \over {\dot a}^3 }\partial_t
\left({ {\dot a}\over {a} }  \right )~, \nonumber \\
Y^{i}~&=&~A~x^i~, \nonumber \\
 Y^5~&=&~A~\left ({r^2\over 4}~-~1~-~{1\over 4 {\dot a}^2}   \right )
-{1\over 2} \int dt ~{ {a}^2 \over {\dot a}^3 }\partial_t
\left({ {\dot a}\over {a}  }  \right )~,
\end{eqnarray}
where $r^2=\eta_{ij}x^ix^j$ and $\eta_{ij}={\rm diag}(1,1,1)$,
the metric takes the form:
\beq
ds^2~=~-(dY^0)^2~+~(dY^1)^2~+~(dY^2)^2~+~(dY^3)^2~+~(dY^5)^2~.
\label{mink}
\eeq
The brane itself in this coordinate system transforms into the
following boundary conditions:
\beq
-(Y^0)^2~+~(Y^1)^2~+~(Y^2)^2~+~(Y^3)^2~+~(Y^5)^2~=~
{ 1\over H_0^2}~, \nonumber \\
Y^0(t, y=0)~>~Y^5(t, y=0)~.
\label{boundary}
\eeq
Therefore, the space to the right of the brane
is transformed to  Minkowski space with the
boundary conditions (\ref {boundary}). 

On this space the S-matrix could be defined as there are asymptotic
{\it in} and {\it out} states of free particles.
The same procedure can be applied to  the metric on  the left of the
brane. However, the brane space-time being de Sitter, one encounters the same 
 problems to define in and out states for scattering products 
localized on the brane. This is true as long as one neglects 
dissipation discussed in section 7, due to which the whole space-time 
will asymptote to Minkowski space-time for which the mentioned problems 
do not persist.

Summarizing, the models with infinite-volume extra dimensions
might be a useful ground for describing an accelerating Universe
within String Theory. In addition we point out that
these models  allow to preserve bulk
supersymmetry even if SUSY is broken on the brane
\cite {DGP0,Wittencc}. Further cosmological studies of these 
models can be found in Refs. \cite {Cedric2}, \cite {Cedric3},
\cite {Arthur1}, \cite {Arthur2}, \cite {others}.

\vspace{0.2in} {\it 6. Massive gravity and perturbation theory}

It has been known for 
some time \cite {Vainshtein} that perturbation theory in massive 
gravity  breaks down at a scale that is parametrically lower than 
an ultraviolet cutoff of the theory.
This breakdown can be traced to  nonlinear graviton self-interaction diagrams
\cite {disc}, and can be interpreted as strong interaction 
of longitudinal polarizations of a massive graviton 
\cite {AGS}. A 
simple way to see the breakdown of perturbation theory is
to look at the tree-level trilinear graviton vertex diagram. 
In the nonlinear level one has  two
extra propagators which could provide a singularity in the graviton mass 
$m_g$ up to  $1/m_g^8$.
Two leading terms $1/m_g^8$ and $1/m_g^6$ do not contribute
so the result contains only the $1/m_g^4$ singularity \cite {disc}.
This leads to breakdown of perturbation theory, and for a
Schwarzschield source of mass $M$ the breakdown  happens at
a scale $\Lambda_m\sim m_g/ (Mm_g/\mpl^2)^{1/5}$ \cite {Vainshtein},
\cite {disc}. As we mentioned above, this can also be understood in terms of
interactions of longitudinal polarizations of a massive graviton
becoming strong \cite {AGS}.  For pure gravitational sector 
itself, that can be thought of as a source with $M=\mpl$,
the corresponding scale $\Lambda_m$  reduces to $m_g/ (m_g/\mpl)^{1/5}$
\cite {AGS}. Using the freedom in addition higher nonlinear
terms this scale can be made only as big as $m_g/ (m_g/\mpl)^{1/3}$
\cite {AGS}.

In Ref. \cite {disc} it has been shown that similar non-linear 
diagrams lead to the precocious breakdown of perturbation theory in 
the model of Ref. \cite {DGP} already at the tree-level.
However, this breakdown was argued to be an artifact of using 
perturbative expansion in $G_N$ which  is ill-defined in that case.
Moreover, it was argued in Ref. \cite {disc} that the 
{\it re-summation} of tree-level diagrams should lead to consistent results. 
This was confirmed  by comparing a number of exact 
solutions of the model of Ref. \cite {DGP} with their perturbative 
counterparts, showing that the perturbative 
results do not reproduce correctly  the results of exact calculations.
Therefore, as long as the classical theory is concerned, 
the model of Ref. \cite {DGP} has no strong coupling problem  
when  it is treated  with consistent methods. 

However, recently it was argued in Refs. \cite {Luty} 
and \cite {Rubakov} that the strong coupling problem could come  
back in loops -- the theory becoming strongly 
interacting  at the quantum level.  
The question of quantum loops, however, is very subtle in 
the present context for the following
reason: there is a connection between the ultraviolet (UV) and 
infrared (IR) physics in this model and the ultraviolet 
completion of the  model is not know.
Hence, discussing quantum loops  at low-energies in a theory with UV/IR 
connection without knowing the UV physics might lead 
to ambiguous results. In particular, choice of a given low-energy prescription
in the loops could lead to implicit assumptions about  the UV theory because 
of the UV/IR connection. In this regard, 
the question whether the precocious breakdown of perturbation 
theory in the loops is an artifact of a low-energy method used 
or it is a fundamental drawback of the theory, remains open, to the 
authors knowledge. This issue will be discussed in detail elsewhere.

\vspace{0.4cm}

{\bf Acknowledgments}
\vspace{0.1cm} \\

I am grateful to the organizers for warm hospitality and 
for creating very productive and enjoyable atmosphere.


\end{document}